%% file: 0_neurips_2019.tex
\documentclass{article}

\PassOptionsToPackage{numbers, compress}{natbib}

\usepackage[preprint]{neurips_2019}

\usepackage[utf8]{inputenc} 
\usepackage[T1]{fontenc}    
\usepackage{hyperref}       
\usepackage{url}            
\usepackage{booktabs}       
\usepackage{amsfonts}       
\usepackage{nicefrac}       
\usepackage{amssymb}
\usepackage{pifont}
\usepackage{microtype}      
\usepackage{comment}
\usepackage{xcolor}
\usepackage{amsmath}
\usepackage{caption}
\usepackage{graphicx}
\usepackage{multirow}
\usepackage{tikz}
\usepackage{subcaption}
\usepackage{wrapfig}
\usepackage{algorithm}
\usepackage[noend]{algpseudocode}
\usepackage[labelfont=bf]{caption}
\usepackage{color}
\usepackage{xfrac}
\usepackage{listings}
\usepackage{parcolumns}
\usepackage[title]{appendix}
\usepackage[para]{footmisc}

\graphicspath{{figs/}} 

\newcommand{\sys}{Coda} 
\newcommand{\sysacc}{82} 
\newcommand{\baseacc}{12} 
\newcommand{\marginacc}{70} 

\newcommand{\cmark}{\text{\ding{51}}}%
\newcommand{\xmark}{\text{\ding{55}}}%
\newcommand{\code}[1]{\texttt{\textbf{#1}}}

\title{A Neural-based Program Decompiler}

\author{%
  Cheng Fu, Huili Chen, Haolan Liu \\
    UC San Diego\\
  \texttt{\{cfu,huc044,hal022\}@ucsd.edu} \\
   \And
    Xinyun Chen \\
    UC Berkeley\\
    \texttt{xinyun.chen@berkeley.edu} \\
   \And
    Yuandong Tian \\
    Facebook\\
    \texttt{yuandong@fb.com} \\
   \And
    Farinaz Koushanfar, Jishen Zhao \\
    UC San Diego\\
    \texttt{\{farinaz,jzhao\}@ucsd.edu} \\   
}

\begin{document}

\maketitle
\vspace{-2em}
\begin{abstract}
    Reverse engineering of binary executables is a critical 
    problem in the computer security domain. On the one hand, malicious parties may recover interpretable source codes from the software products to gain commercial advantages. On the other hand, binary decompilation can be leveraged for code vulnerability analysis and malware detection. However, efficient binary decompilation is challenging. Conventional decompilers have the following major limitations: (i) they are only applicable to specific source-target language pair, hence incurs undesired development cost for new language tasks; (ii) their output high-level code cannot effectively preserve the correct functionality of the input binary; (iii) their output program does not capture the semantics of {the input} and the reversed program is hard to interpret.
    To address the above problems, we propose \sys{}\footnote{\textit{\sys{}} is the abbreviation for \textit{CodeAttack}.}, the first end-to-end neural-based framework for code decompilation. 
    \sys{} {decomposes the decompilation task into of two key phases}: First, \sys{} employs an instruction type-aware encoder and a tree decoder for generating an abstract syntax tree (AST) with attention feeding during the code sketch generation stage. Second, \sys{} then updates the code sketch using an iterative \textit{error correction machine} guided by an \textit{ensembled neural error predictor}. {By finding a good approximate candidate and then fixing it towards perfect, \sys{} achieves superior performance compared to baseline approaches.}
    We assess \sys{}'s performance with extensive experiments on various benchmarks. Evaluation results show that \sys{} achieves an average of \sysacc{}\% program recovery accuracy on unseen binary samples, where the state-of-the-art decompilers yield 0\% accuracy. Furthermore, \sys{} outperforms the sequence-to-sequence model with attention by a margin of \marginacc{}\% program accuracy. {Our work reveals the vulnerability of binary executables and imposes a new threat to the protection of Intellectual Property (IP) for software development.}
\end{abstract}

\input{1_intro}

\input{2_problem}
\input{3_globalFlow}

\input{4_method}
\input{5_eval}
\input{6_relatedWork}

\vspace{-1.2em}
\section{Conclusions and Future Work}
\vspace{-1em}

In this paper, we present \sys{}, the first neural-based decompilation framework that is corroborated to preserve both the semantics and the functionality of the high-level program. \sys{} consists of two key phases for program RE. First, \sys{} generates the high-level program with a high token accuracy leveraging an instruction-aware encoder and an AST decoder network architecture with attention. Next, \sys{} iterative correct errors with the guidance of the ensembled EP which further improves \sys{}'s token and program accuracy. 
Extensive experiments on various benchmarks corroborate that \sys{} outperforms the Seq2Seq model and traditional decompilers by a large margin. 
We believe that our work is a milestone for program security and decompilation.

Meanwhile, we observe that several challenges remain in our current framework that can be addressed in the future work: (i) There are no explicit ending symbols in decompilation task. Future research can tackle this issue to RE large-size binary file. (ii) Previous works on the identification of pointer structures / data type can be incorporated into \sys{} to RE more complicated applications. 


\bibliographystyle{IEEEtran}
\bibliography{ref}
\appendix
\input{7_appendix}









\end{document}

%% file: 1_intro.tex
\vspace{-1.5em}
\section{Introduction} \label{sec:intro}
\vspace{-0.8em}

Decompilation is the process of translating a binary executable to the corresponding high-level code. This technique has been widely used in various security applications, such as malware analysis and vulnerable software patching~\cite{182795,7546501}.
Malicious attackers can also use decompilers to reverse engineer (RE) the commercial off-the-shelf (COTS) software products and reproduce it for illegal usage~\cite{lee2011tie}.
Decompilation is a challenging task since the semantics in the high-level programming language (PL) is {obliterated} during compilation. 
Existing decompilers are language-specific and incur tremendous engineering overhead when extending to new PLs. Furthermore, they fail to preserve the semantic information in the target high-level PL ({see Appendix F}), thus the output is hard to interpret.

It is intuitive that decompilation can be formulated as a general program translation task. Recently, an increasing number of neural network (NN)-based approaches have been proposed to tackle natural language translation problems. For instance, sequence-to-sequence (Seq2Seq) based models achieve the state-of-the-art performance on program translation~\cite{Nguyen,7372046}. We identify three main subroutines in code decompilation: (i) learning control dependency from the connections between basic blocks in the low-level code; (ii) learning data dependency from the register usage and memory access; (iii) learning the grammar of the target PL. 
A straightforward neural-based solution is to use an autoencoder-decoder for translating the low-level program to the high-level code. {Katz et al.~\cite{katz2018using} present} a Recurrent Neural Network (RNN)-based method for decompilation. However, we observe that the naive Seq2Seq models are not suitable for decompilation due to the following reasons. 
First, the inputs to the decompiler are structured low-level statements\footnote{We refer each line of the code as a statement.} that have different construction formats (e.g., number and type of operands). Processing the program as sequence inputs ignores the statement boundaries, thus breaks the modular property of the input program. 
Second, the output program of the Seq2Seq model has a lower probability of capturing the grammar of the target PL since the output is sequentially generated without explicit boundaries. Third, the three subroutines mentioned above are entangled together in the Seq2Seq model, making the learning process hard.   

In this work, we propose \sys{}, a neural program decompilation framework that resolves the above limitations. The requirement to yield a perfect program recovery is very hard to fulfill using a single autoencoder, especially for long programs. As such, \sys{} decomposes decompilation into two sequential phases: \textit{code sketch generation} and iterative \textit{error correction}. By finding a good {approximate} program and then iteratively updating it towards the perfect solution using dynamic information, \sys{} engenders superior performance compared to the single-phase decompilers.

\noindent {\tikz\draw[black,fill=black] (-0.5em,-0.5em) rectangle (-0.2em,-0.2em);} \textbf{Phase 1.} \sys{} uses an instruction type-aware encoder and a abstract syntax tree (AST) decoder for translating the input binary into the target PL. Our encoder deploys separate RNNs for different types of statements, thus the statement boundaries are preserved. Furthermore, the control and data dependency in the input program are translated to the connections between the hidden states of corresponding RNNs. The output from the AST decoder maintains {the dependency constraints and statement boundaries} using terminal nodes, which facilitates learning the grammar of the target PL. 
\noindent {\tikz\draw[black,fill=black] (-0.5em,-0.5em) rectangle (-0.2em,-0.2em);} \textbf{Phase 2.} In this stage, \sys{} employs an RNN-based error predictor (EP) to identify potential prediction mistakes in the output program from Phase 1. Ensembling method can be used to boost the performance of the error prediction. The EP is used to guide the iterative correction of the output program. Unlike traditional decompilers which utilize only the syntax information from the input, \sys{} leverages the Levenshtein edit distance (LD)~\cite{hyyro2001explaining} between the compilation of the updated output program and the ground-truth low-level code to prevent false alarms induced by the EP.  


Empowered by the two-phase design, \sys{} achieves an average program accuracy of \sysacc{}\% on various benchmarks. While the Seq2Seq model with attention and the commercial decompilers yield \baseacc{}\% and 0\% accuracy, respectively. We demonstrate that \sys{}'s output preserves both the functionalities and the semantics.    
In summary, this paper makes the following contributions:

\noindent {\tikz\draw[black,fill=black] (-0.5em,-0.5em) circle (-0.15em);} Presenting the first neural-based decompilation framework that maintains both semantics and functionalities of the target high-level code. 
    \vspace{-0.5em}

\noindent {\tikz\draw[black,fill=black] (-0.5em,-0.5em) circle (-0.15em);} Incorporating various design {principles} to facilitate the decompilation task. More specifically, \sys{} deploys instruction type-aware encoder, AST tree decoder, attention feeding, iterative error correction that leverages both the static syntax and dynamic information.
    \vspace{-0.5em}

\noindent {\tikz\draw[black,fill=black] (-0.5em,-0.5em) circle (-0.15em);} Enabling an efficient \textit{end-to-end} decompiler design. \sys{} can be easily generalized to RE executable in a different hardware instruction set architecture (ISA) or PL with negligible engineering overhead.
    \vspace{-0.5em}

\noindent {\tikz\draw[black,fill=black] (-0.5em,-0.5em) circle (-0.15em);} Corroborating \sys{}'s general applicability and superior performance on various synthetic benchmarks and real-world applications.

 

This is the first paper that provides a holistic and effective solution to the decompilation problem using deep learning. Our work sheds new light on the vulnerability of open sourcing binary executables without any protection. More specifically, we show that the attacker can recover interpretable high-level code with correct functionality from the binary file, which imposes a significant threat on the Intellectual Property (IP) of the program developer. 




%% file: 2_problem.tex
\vspace{-1.2em}
\section{Program Decompilation Problem}  \label{sec:prob}
\vspace{-0.8em}
We introduce the background of low-level code construction and potential challenges in decompilation in Section~\ref{sec:prelim}. The formal definition of code decompilation and our threat model is given in Section~\ref{sec:def} and Section~\ref{sec:threat}, respectively. 

\vspace{-1em}
\subsection{Preliminaries and challenges}  \label{sec:prelim}
\vspace{-0.8em}

\begin{figure}[ht!]
    \centering
    \includegraphics[width=1  \textwidth]{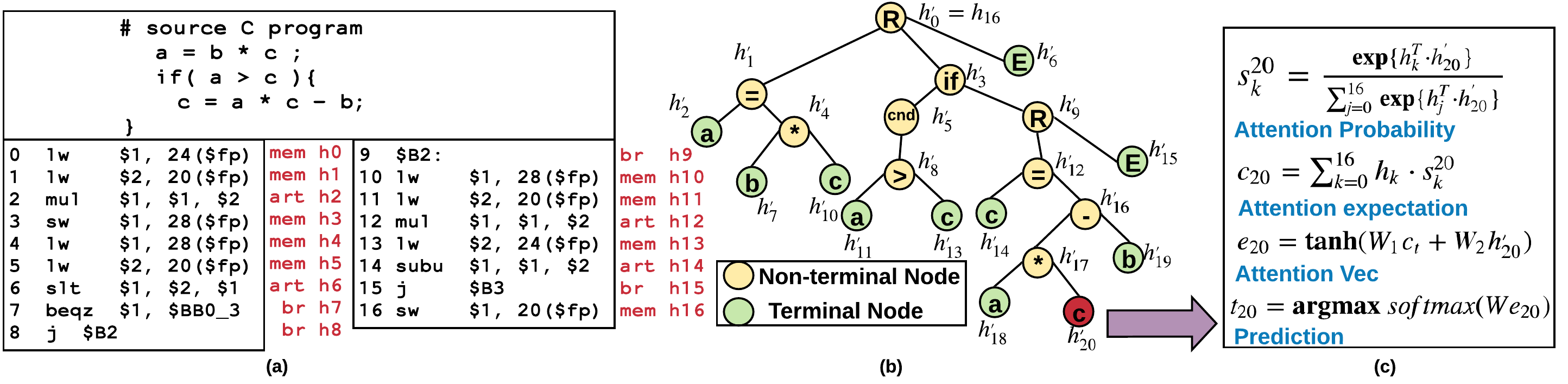} 
    \vspace{-1.6em}
    \caption{(a) Example low-level assembly code snippet and its corresponding high-level C program. The red line indicates the instruction type and its encoded hidden state. (b) The expanding nodes from the AST decoder. (c) The red node is an example of how the prediction is computed.}
    \vspace{-1.2em}
    \label{fig:decompile_example}
\end{figure}

Contemporary software development consists of the following steps: high-level programming, code compilation, deploying the obtained binary files to the pertinent hardware.    
During the execution, a sequence of instructions is carried out on the hardware. 
There are three main instruction types, namely, \textit{memory}, \textit{arithmetic}, and \textit{branch} operations. Different instruction types feature different \textit{instruction fields}, indicating various types and numbers of operands. 
Figure~\ref{fig:decompile_example} (a) shows an example of the high-level code snippet and the corresponding low-level code. 
\code{Line 0} is a memory instruction which fetched a word into register \code{\$1} from the memory address computed from register \code{\$fp} and \code{24}. 
\code{Line 3} is an arithmetic operation which multiplies the value stored in \code{\$1} and \code{\$2}. 
\code{Line 8} refers to an unconditional branch requiring one operand as opposed to three. {Note that \code{lw,mul,j} are the \textit{opcodes} of the instructions.}
Program decompilation is challenging since there are two types of dependencies existing in the low-level program that shall be preserved by the decompiler. 

\vspace{-0.5em}
\noindent {\tikz\draw[black,fill=black] (-0.5em,-0.5em) rectangle (-0.2em,-0.2em);} \textbf{Intra-statement dependency.} Each instruction has a strict structure restriction on the operands as required by the grammar of the low-level ISA. 
For example, in the instruction \code{lw~\$2,8(\$fp)}, the first and the third operand represent registers while the second operand is an instant value. 

\vspace{-0.6em}
\noindent {\tikz\draw[black,fill=black] (-0.5em,-0.5em) rectangle (-0.2em,-0.2em);} \textbf{Inter-statement dependency.} Besides the constraints in a single instruction, \textit{control flow} and \textit{data dependency} exist across multiple instructions. For instance, line 2 and line 3 has data dependency since the \code{mul} operation needs to consume the value from the load destination register. 






\vspace{-1.1em}

\vspace{-0.6em}
\subsection{Problem definition}  \label{sec:def}
\vspace{-0.8em}
We define the task of \textit{Program Decompilation} as follows:

\vspace{-0.3em}
\textbf{Problem Decompilation Definition:} \textit{Let ${P}$ denote an arbitrary program in the high-level language and $\Gamma$ denote the compiler. Given the low-level code $\phi = \Gamma({P})$, the mission of decompilation is to develop a decompiler $\Gamma^{-1}$ that satisfies $\Gamma({P}) = \Gamma({P}^{\prime})$ where ${P}^{\prime} = \Gamma^{-1}(\phi)$}. 

\vspace{-0.5em}
We observe that traditional decompilers such as RetDec or Hex-Rays are only targeted to maintain the functionality of the binary code during decompilation. \sys{} is motivated to address the above limitation by recovering a high-level program with both correct functionality and semantics. 
Besides, we identify two types of constraints of the high-level program that can be explored to verify the correctness of the decompiler's output.
 
\vspace{-0.3em}
\textbf{Input-output Behavior Constraint:} \textit{Given a set of input-output pairs $\{(I^k,O^k)\}_{k=1}^{K}$ where $O^k=\phi(I^k)$ is obtained by executing the low-level program $\phi$, the decompiler shall output a program ${P^{\prime}}$ such that ${\phi^{\prime}}(I^k) = O^k$ for every $k\in{1,...,K}$ where $\phi^{\prime}=\Gamma(P^{\prime})$.}

\vspace{-0.3em}
\textbf{Compilation Matching Constraint:} \textit{The ideal LD between the compilation result $\phi^{\prime}$ of the correctly recovered program and the input low-level code $\phi$ is zero under the same compiler configuration.}








\vspace{-1em}
\subsection{Threat Model}  \label{sec:threat}
\vspace{-0.8em}
We assume the attacker has the following information: (i) the \textit{compiler configuration} that is used to generate the input program; (ii) the interface of \textit{static/dynamic libraries } included in the high-level code; (iii) the disassembler for the pertinent hardware.
The above information can be easily obtained using binary analysis techniques in previous work~\cite{bao2014byteweight,rahimian2015bincomp,rosenblum2008learning}. 
Our objective is to RE a high-level program that depicts the correct computation graph (control and data dependency) and preserves semantic information and functionality as the source high-level program. Reconstruction of data types~\cite{lee2011tie}, finding the function entry point in binary~\cite{rosenblum2008learning,190918,bao2014byteweight} or reconstruct meaningful variable names~\cite{Jaffe_static} are different research directions that have been studied in prior works.

%% file: 3_globalFlow.tex
\vspace{-1.4em}
\section{\sys{} Overview}  \label{sec:overview}
\begin{figure}[ht!]
    \centering
    \includegraphics[width=0.9\textwidth]{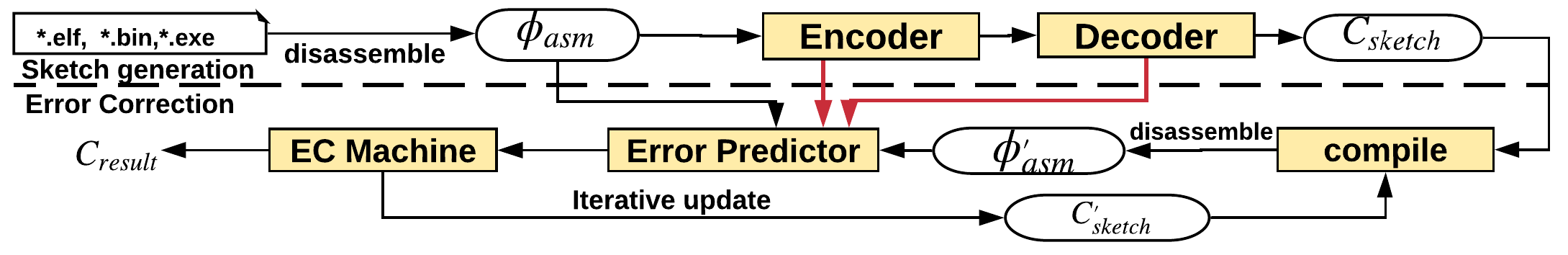} 
    \vspace{-1em}
    \caption{The global flow of \sys{} decompilation. The denoising and tokenization steps are omitted in this figure for simplicity (See Appendix A.1).
    }
    \label{fig:global}
\vspace{-2em}
\end{figure}

\vspace{-1em}
Figure~\ref{fig:global} shows the global flow of \sys{}.
\sys{} framework consists of two key phases: (i) High-level code sketch generation and (ii) iterative error correction. We detail these two phases as follows.

\vspace{-1em}

\vspace{-0.1em}
\subsection{Code Sketch Generation}  \label{sec:overview}
\vspace{-0.7em}
We employ the neural encoder-decoder architecture for generating the sketch code from the low-level program $\phi$.
In this paper, the encoder takes the assembly program generated from the disassembler as the input and output an AST that can be equivalently converted to the high-level program. We discuss the key modules of \sys{}'s code sketch generation below.  

\vspace{-0.3em}
\noindent {\tikz\draw[black,fill=black] (-0.5em,-0.5em) rectangle (-0.2em,-0.2em);} \textbf{Instruction-type Aware Program Encoding.}
\sys{} employs the N-ary Tree-LSTM presented in~\cite{tai2015improved} as the input encoder to handle different instruction types, namely, memory, arithmetic, and branch. 
More specifically, each statement in the input low-level program is fed to the designated LSTM that handles the corresponding instruction type for encoding. 


\vspace{-0.3em}
\noindent {\tikz\draw[black,fill=black] (-0.5em,-0.5em) rectangle (-0.2em,-0.2em);} \textbf{Tree Decoder for AST Generation.} 
We observe that PLs  have more rigorous restrictions on the syntax and semantics compared to natural languages. 
\sys{} opts to use tree decoder for AST generation because of the following advantages: (i) The code statement boundary is naturally preserved by the tree decoder using the terminal node representation. 
(ii) The nodes that are connected in the AST indicate that their corresponding statements in the input program have dependency constraint. Note that the spatial distance for these statements in the program might be large. 
(iii) The error propagation problem during code generation is mitigated using the tree decoder compared to sequential generation.
(iv) AST representation facilitates the verification of syntax restriction. 



\vspace{-0.3em}
\noindent {\tikz\draw[black,fill=black] (-0.5em,-0.5em) rectangle (-0.2em,-0.2em);} \textbf{Attention Feeding.} Our evaluation result shows that code sketch generation is ineffective without attention mechanism (achieving a token accuracy of only {55\%}). 
\sys{} applies parent attention and input instruction attention feeding mechanism~\cite{chen2018tree,dong-lapata-2016-language} that feed the information of the parent node and the input nodes during node expansion performed by the decoder. 

\vspace{-1.0em}
\subsection{Iterative Error Correction}  \label{sec:overview}
\vspace{-0.8em}
The output AST from the code sketch generation phase might contain prediction errors.
As such, we construct an error predictor (EP) and an iterative \textit{error correction machine (EC machine)} as shown in Figure~\ref{fig:global}.  
Specifically, we freeze the autoencoder-decoder from the previous stage and reuse them 
to generate the states of the input nodes $h_k,k = 0,..., K$ and output nodes $h_t, t = 0,..., T$. Here, $K$ and $T$ denote the total number of input states and output states from sketch generation stage. These states ($h_k$ and $h_t$) are fed as the input to the EP. 
Furthermore, \sys{} leverages compiler verification to remove false alarm made by the EP. Note that the input-output behavior of the ground-truth binary executable can also be used as constraints that eliminate false alarms from EPs. To push the performance even further, we ensembled multiple EPs to cover more errors in the decompiled code. 





\vspace{-0.4em}
\noindent {\tikz\draw[black,fill=black] (-0.5em,-0.5em) rectangle (-0.2em,-0.2em);} \textbf{Iterative Error Correction machine.} 
The output of the ensembled EP (containing the location and error type information in the code sketch) is passed to the {EC machine}.  
Note that the {EC machine} prioritizes the potential correction strategies based on the confidence scores obtained from the EP. During each iteration of the error correction process, \sys{} first corrects a single error and validate the resulting high-level code sketch by checking the LD between the compiled code sketch and ground-truth as mentioned above. If the error correction is successful, \sys{} proceeds to the next iteration where EP generates new guidance for the {EC machine}. 





%% file: 4_method.tex
\vspace{-1.4em}
\section{Methodology} \label{sec:method}
\vspace{-1em}
In this section, we detail two key phases of \sys{}'s design as shown in Figure~\ref{fig:global}: autoencoder based code sketch generation, and neural-based iterative error correction. 


\vspace{-1.2em}
\subsection{Autoencoder-based Code Sketch Generation} \label{sec:sketch_gen}
\vspace{-0.8em}

We introduce the main modules of \sys{}'s code sketch generation phase as follows. 

\vspace{-0.3em}
\noindent {\tikz\draw[black,fill=black] (-0.5em,-0.5em) rectangle (-0.2em,-0.2em);} \textbf{Instruction-aware encoding.}
The computation flow of \sys{}'s input program encoding is shown in Equation~(\ref{eq:input_embedding}) where the subscript $n$ refers to $n_{th}$ instruction statements. To capture the intuition of learning the intra and inter-dependency of the instruction statements as discussed in Sec~\ref{sec:prob}, \sys{} employs an N-ary  Tree Encoder~\cite{tai2015improved} which is suitable for encoding task where the children are structured. 
The input states are fed into the N-ary Tree Encoder with a consistent order of the corresponding instruction type. As such, the intra-statement dependency can be effectively captured. 
Particularly, \sys{} designates a specific N-ary encoder for each instruction type, i.e., memory, branch and arithmetic instructions ($LSTM_i$ where $i \in \{mem,br,art\}$). 
Note that in \sys{}'s code encoding process, each {non-terminal} node has at most 4 children, consisting of the embedded states of up to 3 operands in the current instruction ($h^{op}_i, c^{op}_i$ where $i = 0,1,2$), and the context vector of the previous instruction ($h_n, c_n$). \sys{} encodes the input instructions with the maximal number of operands (i.e., 3) and pads short statements with zero states. 
The input $x$ is the embedding of the {instruction opcode} of the current statement.
The basic block header (e.g., $\$B2$ in line 9 of Figure~\ref{fig:decompile_example} (a)) are also handled as branch instructions. 

\vspace{-2em}
\begin{equation}  \label{eq:input_embedding}
(h_{n+1},c_{n+1}) = LSTM_i(([h_n;h^{op}_{0};h^{op}_{1};h^{op}_{2}],[c_n;c^{op}_{0};c^{op}_{1};c^{op}_{2}]),~x),i \in \{mem,br,art\}
\end{equation}
\vspace{-1.7em}


\vspace{-0.3em}
\noindent {\tikz\draw[black,fill=black] (-0.5em,-0.5em) rectangle (-0.2em,-0.2em);} \textbf{Binary Abstract Syntax Tree decoder.} 
The output states of the last instruction in the low-level code is used as the input to \sys{}'s tree decoder for AST generation.
Non-leaf nodes in general AST structures may have multiple children, which complicates the high-level code generation process since the number of children varies for different nodes. 
To address this uncertainty in the decoding stage, \sys{} generates a binary tree in Left-Child Right-Sibling representation which is equivalent to the target AST output. 
As a result, each sub-tree in the AST output has a regulated structure that is consistent with the root. 
We deploy two LSTMs that predict the left and the right child of the current node separately.  
Note that \sys{}'s output AST does not contain the statement ending token as the termination is naturally represented by the terminal nodes. 
For example, a complete statement $a=b*c$ can be recovered from the AST subtree without an explicit ending token as shown in Figure~\ref{fig:decompile_example}. 
The state transition equations of \sys{}'s AST decoder are shown as follows:
\vspace{-0.3em}
\begin{equation}  \label{eq:decoder_left}
(h_L,c_L) = LSTM_L((h,c),[Ho_t;e_t])
\end{equation}

\vspace{-2.0em}
\begin{equation}  \label{eq:decoder_right}
(h_R,c_R) = LSTM_R((h,c),[Ho_t;e_t])
\end{equation}
\vspace{-1.7em}

Here, the subscript $t$ denotes the current expanding node $N_t$ in the output AST. The symbols $o_t$ and $e_t$ indicate the predicted token value and the attention vector (explained later) of the node $N_t$. 
$H$ is a trainable token embedding matrix with dimension $d \times V$, where $d$ and $V$ are the embedding dimension and the vocabulary size of the high-level PL, respectively.

\vspace{-0.3em}
\noindent {\tikz\draw[black,fill=black] (-0.5em,-0.5em) rectangle (-0.2em,-0.2em);} \textbf{Input Instruction and Parent Attention Feeding.} \label{sec:attention_method} To make better use of the information encoded from the input program and the parent context of the current expanding node, we employ instruction and parent attention feeding during AST decoding~\cite{mccann2017learned, chen2018tree}. 
Intuitively, predicting the current node while leveraging the relevant information from the input instructions and the node's parent provides a richer context for high-level code generation.   
Parent attention feeding is performed using Equation~(\ref{eq:decoder_left}) and~(\ref{eq:decoder_right}) during the state transition of the AST decoder. As for input instruction attention feeding, we first compute the probability that a node $N_k$ in the input program corresponds to the expanding node $N_t$ as shown in Equation~(\ref{eq:prob_cond}). \sys{}'s input instruction attention is obtained from the expectation value of the hidden states of all nodes in the input program as shown in Equation~(\ref{eq:instruc_attn}).

\vspace{-1.2em}
\begin{equation}  \label{eq:prob_cond}
s_k^t = P(N_k | N_t) \propto \textbf{exp}\{h_k^T \cdot h_t\}
\end{equation}

\vspace{-1.7em}
\begin{equation}  \label{eq:instruc_attn}
c_t = \mathbb{E}[h_k | N_t] = \sum_{k=0}^{K} h_{k} \cdot s_k^t
\end{equation}
\vspace{-1.4em}

$c_t$ is then incorporated into the hidden state of the current node $h_t$ using Equation~(\ref{eq:e_t}) where $W_1$ and $W_2$ are two trainable matrices with dimension $d \times d$, resulting in the attention vector $e_t$ of the current node. The final prediction output $o_t$ of the current expanding node is then be computed from the linear mapping of $e_t$ as shown in Equation~(\ref{eq:o_t}). $W_{out}$ is a trainable matrix of size $V \times d$.

\vspace{-1.2em}
\begin{equation}  \label{eq:e_t}
e_t = \textbf{tanh}(W_1c_t + W_2h_t)
\end{equation}

\vspace{-1.7em}
\begin{equation}  \label{eq:o_t}
o_t = \textbf{argmax} ~ softmax({ W_{out}} e_t)
\end{equation}
\vspace{-2.5em} 
\subsection{Neural-based Iterative Error Correction}   \label{sec:iter_EC}

 
\vspace{-0.8em}
We propose iterative Error Correction as the second phase of \sys{} framework to further improve the quality of decompilation as discussed in Sec.~\ref{sec:overview}.  
There are two key modules in this stage: an ensembled neural EP and an Iterative \textit{EC machine}. 
We characterize possible errors in \sys{}'s code sketch into three types: 
(i) Nodes in the AST may be mispredicted to other tokens. For example, the `\code{while}' might be misclassified into `\code{if}' token. 
(ii) A redundant line of code. 
(iii) A Missing line of code. 
For error (i), the EP shall output the correct token value to guide the {EC machine} for updating the node.
For error (ii) and (iii), the {EC machine} removes/randomly adds a non-terminal node with leaf children in the predicted error location, thus converting the error type into a misprediction error (i). (See Appendix B for details)
Equation~(\ref{eq:EP_hidden}) shows the hidden state transition of \sys{}'s EP. We deploy the fixed autoencoder from phase 1 followed by gated recurrent units (GRUs) with attention as the EP's architecture. {Given the ground-truth input ($\phi$) and the compiled code sketch ($\phi^{\prime}$), the EP returns the error status (`0/1') and the error types for each node in the output AST.} 
The input to the GRU consists of two parts: (i) the hidden state of the parent node ($h_{t-1}^{EP}$); and (ii) the concatenation of the hidden states (denoted by $h_t^{\phi}$ and $h_t^{\phi^{\prime}}$) obtained by forwarding $\phi$ and $\phi^{\prime}$ to the autoencoder. 
\vspace{-0.3em}
\begin{equation}  \label{eq:EP_hidden}
h_t^{EP} = GRU(h_{t-1}^{EP},[h_t^{\phi};{{h_t^{\phi^{\prime}}}}])
\end{equation}
\vspace{-1.8em}

The attention layer in EP following the mechanism discussed in Sec.~\ref{sec:attention_method}. Particularly, the input to the attention layer $h_t$ in Equation~\ref{eq:prob_cond} is now replaced by the hidden state $h_{EP}$ of the current node.
The state of source input ($h_k$) is substituted with the combination of the encoded states $h_k^{\phi}$ and $h_k^{\phi^{\prime}}$. Furthermore, \sys{} ensembles multiple EPs to cover larger error space. 
The correction suggestion provisioned by the EP is accepted if and only if the LD between the golden low-level code and the compilation of the updated code sketch does not increase. The workflow of \sys{}'s iterative \textit{EC machine} is shown in Algorithm~\ref{alg:Error_correction_step}. The detail of the function $FSM\_Error\_Correct$ in line 9 is presented in Appendix B.

\algnewcommand\algorithmicinput{\textbf{INPUT:}}
\algnewcommand\INPUT{\item[\algorithmicinput]}
\algnewcommand\algorithmicoutput{\textbf{OUTPUT:}}
\algnewcommand\OUTPUT{\item[\algorithmicoutput]}

\vspace{-0.8em}
\setlength{\belowcaptionskip}{-3em}
\begin{algorithm}[ht!]
\caption{Workflow of iterative EC Machine. }
\label{alg:Error_correction_step}
\begin{algorithmic}[1]
\INPUT \textbf{$N_{EP}$ Ensembled Error Predictors ${EP}$; Source assembly $\phi$; Decompiled Sketch program ${P}^{\prime}$; Compiler $\Gamma$; Maximum iterations $S_{max}$ and steps in each iteration $c_{max}$; }

\vspace{0.5em}
\OUTPUT \textbf{Error corrected program ${P}_f^{\prime}$. }
\vspace{0.2em}
\State $s_i \gets 0 $
\State \textbf{while} $s_i < S_{max}$ \textit{do} 
\State \indent $Q \gets []$,$~\phi^{\prime} = \Gamma({P}^{\prime})$, $ ~\Delta^{\prime} \gets Edit\_loss(\phi,\Gamma({P}^{\prime}))$
\State \indent \textbf{if} $\Delta^{\prime} = 0 $ \textbf{then} \textbf{break}
\State \indent $Q \gets EP_i({P}^{\prime})$ for $i$ = 1,...,$N_{EP}$ \indent// Attach all the detected error to queue $Q$
\State \indent  $\widetilde Q \gets Prob\_sort(Q,c_{max})$  \indent// Rank $Q$ using output probabilities, keep $c_{max}$ results. 
\State \indent \textbf{while} $\widetilde Q$ is not empty \textbf{do}
\State \indent \indent $err,node \gets \widetilde Q.pop()$
\State \indent \indent ${P}_t^{\prime} \gets FSM\_Error\_Correct({P^{\prime}}, err,node)$ \indent // correct the error in the program
\State \indent \indent $\Delta = \Delta^{\prime} - Edit\_loss(\phi,\Gamma({P}_t^{\prime}))$
\State \indent \indent \textbf{if} $\Delta \geq 0 $ \textbf{then}
\State \indent \indent \indent ${P}^{\prime} \gets {P}_t^{\prime} $

\State
\textbf{Return:}  ${P}_f^{\prime}\gets {P^{\prime}}$
\end{algorithmic}
\end{algorithm}
\setlength{\textfloatsep}{-0.1em}

%% file: 5_eval.tex
\vspace{-1.7em}
\section{Evaluation}  \label{sec:eval}
\vspace{-1em}

\subsection{Experimental Setup}  \label{sec:Dataset}
\vspace{-0.8em}
We assess the performance of \sys{} on various synthetic benchmarks with different difficulty levels and real-world applications as summarized in Table~\ref{tab:token_accuracy} (Stage 1) and Table~\ref{tab:program_accuracy} (Stage 2).
{Given the binary executable as the input, we use an open-source disassembler~\cite{mips_disassemble,REDasm} for MIPS~\cite{Kane:1988:MRA:59923} and x86-64~\cite{Intel_x64} architecture to generate the corresponding assembly code that is fed to \sys{}. }

\vspace{-0.3em}
\noindent {\tikz\draw[black,fill=black] (-0.5em,-0.5em) rectangle (-0.2em,-0.2em);} \noindent \textbf{Benchmarks.} We describe the four main tasks in our evaluation as follows. \label{sec:benchmark}
\vspace{-0.3em}

\textbf{(i) Karel.} Karel~\cite{Karel} is a C-based library that can be used to control the movement of a robot in a 2D grid and modify the status of the environment.  
The assembly description of Karel programs has only callback functions (no arguments) and global control flags as {shown in Appendix E}. 
As such, Karel is suitable to evaluate \sys{}'s capability of recovering the control flow graphs (CFGs) of the source code (see possible CFGs in Appendix A.2). 

\vspace{-0.4em}
\textbf{(ii) Math Library (Math).} We generate synthetic benchmarks using \code{math.h} library~\cite{Math} to assess \sys{}'s performance for recovering both data and control dependencies. 
\vspace{-0.4em}

\textbf{(iii) Normal Expression (NE).}
Common operations such as $"+,-,\ast,\backslash,\| ,\gg,\&,==,\wedge"$ are the main components of NEs in high-level PL. 
We observe that reconstructing normal expressions is more difficult compared to function calls since the former one has less explicit structures. 
\vspace{-0.4em}

\textbf{(iv) Composition of Functional Calls and Normal Expressions (Math+NE).} We also construct synthetic benchmarks consisting of both NEs and library functions calls. The dataset is constructed by replacing the variables in NE with the return value of a random math function (see Appendix E).
\vspace{-0.4em}

\textbf{(v) Real-world implementations.} We test \sys{}'s performance on real-world projects: (1) neural network construction programs in pytorch C++ API~\cite{pytorch_cpp} (2) Hacker's Delight loop-free programs~\cite{warren2013hacker} provided in~\cite{Schkufza:2013}. These programs are used for encoding complex algorithms as small loop-free sequences of bit manipulating instructions.


\vspace{-0.4em}
\noindent {\tikz\draw[black,fill=black] (-0.5em,-0.5em) rectangle (-0.2em,-0.2em);} \noindent \textbf{Training Data Generation.} 
To build the training dataset for stage 1, we randomly generate 50,000 pairs of high-level programs with the corresponding assembly code for each task. The program is compiled using \code{clang} with configuration \code{-0O} which disables all optimizations. The subscript $S$ and $L$ in Table~\ref{tab:token_accuracy} denotes short and long programs with an average length of 15 and 30, respectively. The tree representation of each statement in the high-level code has a maximum depth of 3. 

The training dataset for the error correction stage is constructed by injecting various types of errors into the high-level code. 
In our experiments, we inject $10\sim20\%$ token errors whose locations are sampled from a uniform random distribution. 
To address the class imbalance problem during EP training, we mask $35\%$ of the tokens with error status `0' (i.e., no error occurs) when computing the loss. 
Detailed statistics and examples of the dataset can be found in {Appendix A.2 and E}. 





\vspace{-0.5em}
\noindent {\tikz\draw[black,fill=black] (-0.5em,-0.5em) rectangle (-0.2em,-0.2em);} \noindent \textbf{Metrics.} We evaluate the performance of the \sys{} using two main metrics: \textit{token accuracy} and \textit{program accuracy}. Token accuracy is defined as the percentage of the predicted tokens in high-level PL that match with the ground-truth ones. Program accuracy is defined as the ratio between the number of predicted programs with 100\% token accuracy and the number of total recovered programs. 

\vspace{-1.2em}
\subsection{Results}   \label{sec:results}
\vspace{-0.8em}

\noindent \textbf{Performance of Sketch Generation.} 
\sys{} yields the highest token accuracy across all benchmarks (96.8\% on average) compared to all the other methods as shown in Table~\ref{tab:token_accuracy}. The \code{NE} task appears to be the hardest one while \sys{} still engenders 10.1\% and 80.9\% margin over a naive \textit{Seq2Seq} model with and without attention, respectively. 
More importantly, the result demonstrates that our \textit{Inst2AST+Attn} method is more tolerant of the growth of the program length compared to the \textit{Seq2Seq+Attn} baseline. 
We hypothesize that this is because \textit{Inst2AST} with attention focuses on the states of each instruction as a whole instead of every input token. As such, it is less sensitive to the growth of assembly token length. Note that \sys{} achieves higher token accuracy on \code{Math+NE} benchmarks compared to \code{NE} ones. This is due to the fact that the assembly description of function calls has a prologue of argument preparation~\cite{190918} that is easy to identify than \code{NE} which directly operates on variables. 

We observe that the majority of the token errors are misprediction in the sketch, especially when the program size is large. {Besides, the sketch may have missing or repetition statements.} The imperfection of code sketch generation motivates the design of the error correction stage in \sys{}. 
\begin{table}[]
\centering
\caption{Token accuracy (\%) comparison between \sys{} and alternative methods for code generation. Columns 1-2 denotes the \textit{Seq2Seq} baseline. The last two columns denote the instruction-aware encoding (Inst) and AST decoding (AST) methods of \sys{} with and without attention (Attn) mechanism. The combination of a sequence-based model with Inst or AST is shown in Columns 3-4.} 
\label{tab:token_accuracy}
\scalebox{0.85}{
\begin{tabular}{|ccccccc|}
\hline
Benchmarks   & Seq2Seq & Seq2Seq+Attn  & Seq2AST+Attn & Inst2seq+Attn & Inst2AST & Inst2AST+Attn \\ \hline
Karel$_S$        & 51.61       & 97.13  & 99.81&  98.83& 74.80     & \textbf{99.89}\\ \hline
Math$_S$         & 23.12       & 94.85    & 99.12 & 96.20& 56.29    & \textbf{99.72}                 \\ \hline
NE$_S$           & 18.72       & 87.36          & 90.45& 88.48 & 55.59       & \textbf{94.66}                  \\ \hline
(Math+NE)$_S$      & 14.14       & 87.86         & 91.98 &  89.67 & 56.62      & \textbf{97.90}                   \\ \hline
Karel$_L$     & 33.54       & 94.42         & 98.02 & 98.12 & 64.42            & \textbf{98.56}                 \\ \hline
Math$_L$      & 11.32       & 91.94         & 96.63 & 93.16  & 45.63     & \textbf{98.63}                 \\ \hline
NE$_L$        & 11.02       & 81.80         & 85.92 & 85.97& 46.43             & \textbf{91.92}                 \\ \hline
(Math+NE)$_L$ & 6.09       & 81.56         & 85.32 & 86.16 & 43.77             & \textbf{93.20}                 \\ \hline
\end{tabular}
}
\end{table}



\vspace{-0.3em}
\noindent \textbf{Performance of Error Correction.} Table~\ref{tab:program_accuracy} summarizes the performance of \sys{}'s iterative error correction. We feed the recovered code sketch with imperfect token accuracy generated from stage 1 to the pretrained EP. Recall that the EP consists of the fixed autoencoder from stage 1 and a GRU layer. Here, EPs that reuse the \textit{Seq2Seq+attn} and \textit{Inst2AST+attn} sketch generator are denoted as EP$_{s2s}$ and EP$_{i2a}$, respectively. We set $S_{max}=30$ and $c_{max}=10$ for \textit{EC machine} in Algorithm~\ref{alg:Error_correction_step}.
A single EP achieves 66\% (EP$_{s2s}$) and 69\% (EP$_{i2a}$) accuracy on average across benchmarks for predicting the error type in the sketch programs. Note that we only consider the prediction of the first error due to the iterative nature of the EC machine. 

When ensembling 10 EPs ($N_{EP} = 10$), the detection rate of first error can be enhanced to 84\% and 89\% on average for EP$_{s2s}$ and EP$_{i2a}$, respectively. 
Note that EP$_{i2a}$ achieves a higher accuracy on error prediction across benchmarks compared to  EP$_{s2s}$. That is because the component of the EP$_{i2a}$, i.e., \textit{Inst2AST+attn}, achieves a better token accuracy compared to \textit{Seq2Seq+attn} in EP$_{s2s}$. 

{The ensembled EPs will guide our iterative EC machine as detailed in Algorithm~\ref{alg:Error_correction_step}. \sys{}'s EC machine increases the program accuracy from 12\% to 61\% and from 30\% to 82\% on average for \textit{Seq2Seq+Attn}-based and \textit{Inst2AST+attn}-based code sketch generation, respectively.}
In summary, \sys{}'s best configuration (\textit{Inst2AST+attn} with EC) achieves an average of {\sysacc{}}\% final program accuracy while the \textit{Seq2Seq} model with or without attention approach yields \baseacc{}\% and {0\%}, respectively.

\vspace{-0.4em}
\begin{table}[ht!]
\centering
\caption{(i) First error prediction accuracy with various ensembled number of ensembled EPs. (ii) Program accuracy before and after error correction (EC) when $N_{EP}$=10. Note that $N_{EP}$ stands for the number of ensembled EPs and model refers to the architecture of sketch generation.}
\label{tab:program_accuracy}
\scalebox{0.84}{
\begin{tabular}{|c|c|c|c|c|c|c|c|c|c|c|}
\hline
\multirow{2}{*}{BenchMarks} & \multicolumn{6}{c|}{(i) First Error Detection Rate (EP,$N_{EP}$)} & \multicolumn{2}{c|}{(ii) Befor EC} & \multicolumn{2}{c|}{After EC} \\ \cline{2-11} 
 &  ${s2s}$, 1 & ${i2a}$,1 &  ${s2s}$,5 & ${i2a}$,5 &  ${s2s}$,10 & ${i2a}$,10 & ${s2s}$ & ${i2a}$ & ${s2s}$ & ${i2a}$ \\ \hline
Math$_S$ & 69.6 & 74.1 & 84.9 & 88.5 & 91.4 & 94.2 & 40.1 & 64.8 & 91.2 & \textbf{100.0} \\ \hline
NE$_S$ & 64.2 & 67.6 & 76.0 & 79.2 & 83.5 & 88.7 & 6.6 & 12.2 & 53.0 & \textbf{78.6} \\ \hline
(Math+NE)$_S$ & 65.1 & 67.3 & 78.4 & 84.4 & 83.6 & 90.1 & 3.5 & 43.2 & 63.6 & \textbf{89.2} \\ \hline
Math$_L$ &  65.4 & 71.7 & 80.9 & 83.1 & 87.5 & 91.3 & 21.7 & 51.8 & 83.9 & \textbf{99.5} \\ \hline
NE$_L$ & 60.3 & 64.7 & 71.6 & 76.5 & 78.1 & 84.5 & 0.2 & 2.6 & 33.1 & \textbf{56.4} \\ \hline
(Math+NE)$_L$ & 61.0 & 66.5 & 73.9 & 77.5 & 80.2 & 85.3 & 0.1 & 4.9 & 38.3 & \textbf{67.2} \\ \hline
\end{tabular}
}
\end{table}

\noindent \textbf{Results on Real-world Applications.} We assess \sys{} on two real-world applications: Pytorch C++ API-based~\cite{pytorch_cpp} model architecture construction and bit twiddling hack in Hacker's Delight~\cite{warren2013hacker}.  
The model definition and the bit twiddling task programs consist of a sequence of function calls and a sequence of loop-free normal expressions, respectively. {Examples of these two applications are given in Appendix E.}   
\sys{} achieves 100\% program accuracy across all benchmarks for these two tasks.


\noindent \textbf{Comparison to Previous Works.} We demonstrate that \sys{} outperforms two state-of-the-art decompilers: RetDec~\cite{ret_dec} open-source tool and sequence-to-sequence based decompiler~\cite{katz2018using}. The output from RetDec has a large LD to the ground-truth low-level code after compilation. Furthermore, the high-level program recovered by RetDec fails to preserve the functionality of the input and is hard to interpret. (See {Appendix F} for examples). 
The \textit{Seq2Seq}-based approach proposed in~\cite{katz2018using} takes a sequence of bytes or bits directly from the binary executable as the input. We re-implement their technique and assess its performance on \code{Math+NE} synthetic benchmarks. Empirical results show that their \textit{Seq2Seq}-based method achieves 11\% token accuracy and 0\% program accuracy on average. 


%% file: 6_relatedWork.tex
\vspace{-1.2em}
\section{Related Work} 
\label{sec:related}
\vspace{-0.9em}

\noindent \textbf{Conventional Program Decompilation.} There has been a long line of research on reverse engineering of binary code~\cite{cifuentes1994reverse,emmerik2004using,10.1007/978-3-642-22110-1_37,bao2014byteweight,rosenblum2008learning,brumley2013native,7546501}. 
Many decompilers such as Phoenix~\cite{brumley2013native},  Hex-rays~\cite{hex_ray} and RetDec~\cite{ret_dec} (the most recent one) have been developed. 
Other works, such as TIE~\cite{lee2011tie} or REWARDS~\cite{dolan2011virtuoso}, target at reconstructing the correct variable types, which is different from the objective of \sys{}.
Learning-based methods have been proposed for identifying function entry point~\cite{bao2014byteweight,rosenblum2008learning,190918} for disassembling binary code. These methods are orthogonal to \sys{} and can be integrated into our framework to tackle a wider range of decompilation tasks. To the best of our knowledge, no practical deep learning-based techniques have been proposed for program decompilation. 


\vspace{-0.3em}
\noindent \textbf{Neural Networks for Code Generation.} Neural networks have been used for code generation in prior works~\cite{ling-etal-2016-latent,yin-neubig-2017-syntactic,rabinovich2017abstract,chen2018tree,chen2017towards}. Instead of recovering the high-level program from the corresponding assembly code, these works synthesize the program from the collected input-output pairs~\cite{chen2018executionguided,chen2017towards}, natural language~\cite{yin-neubig-2017-syntactic}, or other domain-specific languages~\cite{chen2018tree,Nguyen}. In~\cite{katz2018using}, they use a sequence-to-sequence model for decompilation with direct Byte-by-byte sequence input which yields a low accuracy as shown in Sec.~\ref{sec:eval}. \sys{} demonstrates the first effective program decompilation framework. 

\vspace{-0.3em}
\noindent \textbf{Neural Networks for Error Correction.} The idea of iteratively fixing errors in the program using neural networks has been proposed~\cite{gupta2017deepfix,piech2015learning,wang2017dynamic}. In~\cite{piech2015learning}, they suggest using GRUs for embedding the execution trace in order to identify bugs in the target program. DeepFix~\cite{gupta2017deepfix} deploys an autoencoder to fix typos that leads to a compilation failure. Note that the error correction stage of \sys{} has a different objective from the above works. More specifically, we use an autoencoder-based error predictor to identify the token errors in output from the code sketch generation stage. 




%% file: 7_appendix.tex

\section{Experiment Setup and Benchmark Details}
\vspace{-0.8em}
We ran our experiments on Amazon EC2 using \code{p3.4xlarge} instance which contains Nvidia Tesla V100 GPUs with 16GB main memory.

\vspace{-0.8em}
\subsection{Denoising Process}
\vspace{-0.8em}
We start the tokenization of the low-level assembly from the beginning of the program function to be reversed, e.g., \code{`func:'} in assembly of our case. All linkers (such as \code{`.cfi*'}), no-ops (\code{`nop'}), brackets and commas are removed.

\vspace{-0.8em}
\subsection{Dataset Statistics}
\vspace{-0.8em}
We present the detailed statistics of the datasets in Table~\ref{tab:statistics_bench} used in Evaluation Section.

\vspace{-1.2em}
\begin{table}[ht!]
\centering
\caption{Statistics of the datasets benchmarks used in Evaluation Section.}
\label{tab:statistics_bench}
\scalebox{0.85}{
\begin{tabular}{|c|c|c|c|c|c|c|}
\hline
 Length (tokens)& Karel$_{S/L}$ & Math$_{S/L}$ & NE$_{S/L}$ & (Math+NE)$_{S/L}$ & Pytorch & Hacker's light \\ \hline
Average output  &  39/76 & 42/89 &  57/111& 72/142 & 36 & 25 \\ \hline
Average input  &  126/247 & 219/423 & 323/485 & 334/637 & 190 & 104  \\ \hline
Minimal variable number & 0 & 3/10 & 3/9 & 3/9 & 3 & 2 \\ \hline
Maximal variable number & 0 & 8/15 & 8/15 & 8/15 & 10 & 16 \\ \hline
branch program & 	$\cmark$ & $\cmark$ & $\cmark$ & $\cmark$ & $\xmark$ & $\xmark$ \\ \hline
\end{tabular} 
}
\end{table}


\vspace{-0.8em}
\subsection{Random Control Flow Graphs Generation}
\vspace{-0.7em}

\begin{figure}[ht!]
    \centering
    \includegraphics[width=\textwidth]{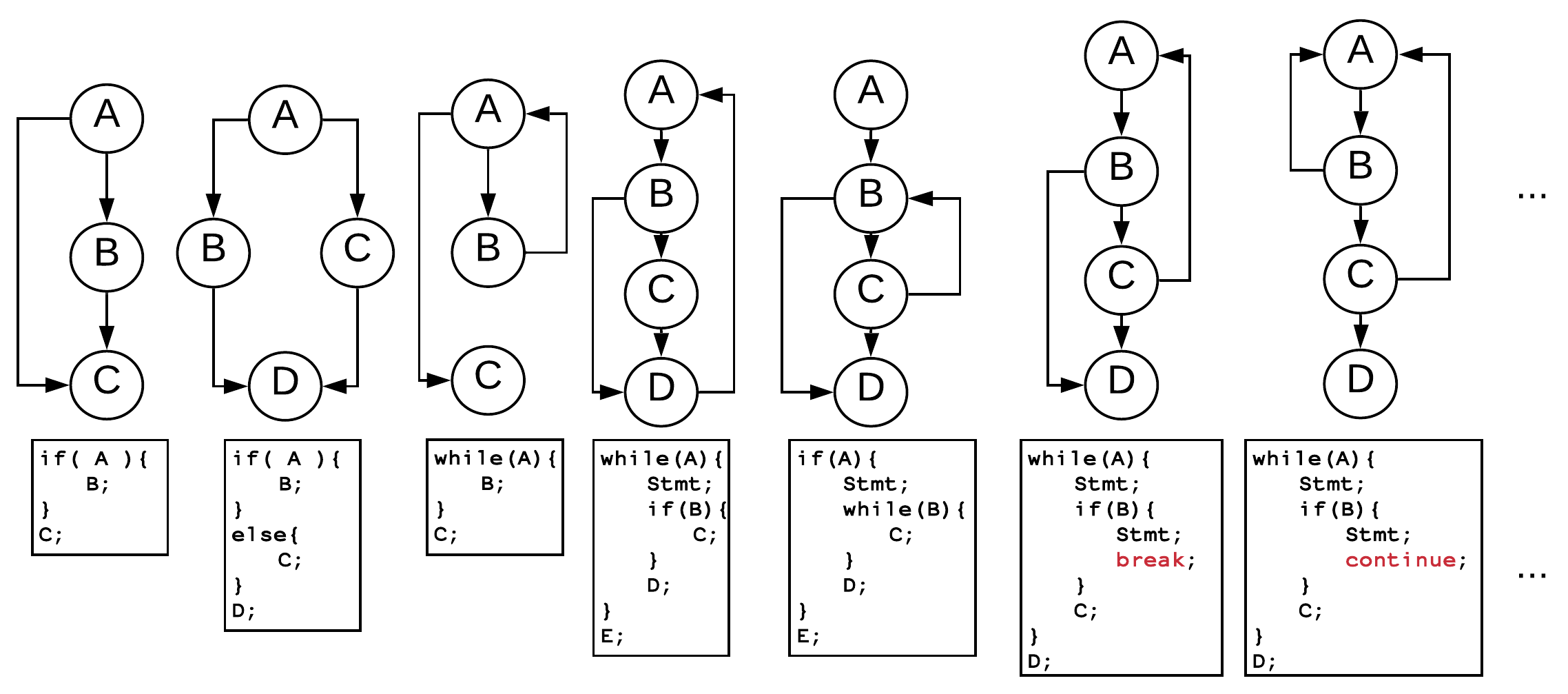} 
    \vspace{-1em}
    \caption{Examples of the possible control flow graph of dataset generation for \sys{} decompilation.}
    \label{fig:cfg_example}
\end{figure}

We evaluate various CFGs in our synthetic benchmarks. More specifically, multiple basic random CFGs (shown in Figure~\ref{fig:cfg_example}) are generated and connected to form the final CFG for the input program. 








\vspace{-1em}
\vspace{2em}

\newpage

\vspace{-1em}
\section{ Details of $FSM\_Error\_Correct$ Function}
\vspace{-1em}

Algorithm~\ref{alg:Correct_error_fsm_2} details the $FSM\_Error\_Correct$ function in Algorithm~\ref{alg:Error_correction_step}. As mentioned in the Method Section, the Error Predictor (EP) outputs three error types, namely, (1) mispredicted tokens, (2) missing lines, and (3) redundant lines. 
For mispredicted error type, the Error Correction machine (EC machine) replaces the node token with the prediction output of the ensembled EP. {As a special case of the mispredicted error where the terminal node is guided to become a non-terminal node, the EC machine adds random children to corresponding non-terminal node.} As for missing errors, the EC machine adds extra a non-terminal node to the root node in the corresponding position, indicating the addition of a random newline to the output program. For the redundant error type, \sys{} removes the particular node from its parent. The structural difference between the mispredicted, missing and redundant errors is that the misprediction error occurs inside a program statement whereas the missing/redundant error indicates a lost/additional statement in the program. \sys{}'s EC machine first transforms the missing/redundant error into the mispredicted error and then proceeds to iteratively fix the error insides the line.

\begin{algorithm}[ht!]
\caption{ Correct Error FSM algorithm. }
\label{alg:Correct_error_fsm_2}

\begin{algorithmic}[1]
\INPUT \textbf{Decompiled program ${P}^{'}$; error type $err$ ; Node id $node$ ; $\perp$ terminal tokens;  three types of error, viz, mispredicted $pred$, missing line $ms$, redundant line $rdt$}

\vspace{0.5em}
\OUTPUT \textbf{Modified program ${P}_t^{'}$. }
\vspace{0.2em}

\State  \textbf{if} $err \in pred$ and $ ~node \in \perp$ and $ ~err \in \perp $  \textbf{then} 
\State  \indent $node.value \gets err$
\State  \textbf{else if} $err \in fp$ and $ ~node \notin \perp$ and $ ~err.value \notin \perp $  \textbf{then} 
\State  \indent $node.value \gets err.value$
\State  \textbf{else if} $err \in pred$ and $ ~node \notin \perp$ and $ ~err.value \in \perp $  \textbf{then} 
\State  \indent $node \gets newNode()$
\State  \indent $node.value \gets err.value $
\State  \textbf{else if} $err \in fp$ and $ ~node \in \perp$ and $ ~err \notin \perp $  \textbf{then} 
\State  \indent $node \gets Add\_Random\_Children(node,\perp)$ \indent 

//assign random terminal children for the node
\State  \indent $node.value \gets err.value$
\State  \textbf{else if} $err \in rdt$ \textbf{then} 
\State  \indent $node.parent.remove(node)$ \indent //remove the node from its parent
\State  \textbf{else if} $err \in ms$ \textbf{then}
\State  \indent $node^{\prime}\gets newNode(not \perp)$
\State  \indent $node^{\prime}\gets Add\_Random\_Children(node^{\prime}, \perp)$ \indent 
\State  \indent $node.parent.Add\_Children\_Behind(node^{\prime},node)$ 
//add random non-terminal node to the root node behind the current node
\State \textbf{endif}
\State
\noindent \textbf{Return:} ${P}_t^{'} \gets {P^{'}}$
\end{algorithmic}
\end{algorithm}

\newpage
\section{Hyper-parameters of Neural Network Models}
\vspace{-1em}

We present the hyper-parameters used by different neural networks in Table~\ref{tab:hyper}. These hyper-parameters are selected to achieve the best accuracy using cross-validation with \textit{grid search}.

\vspace{-1em}
\begin{table}[ht!]
\centering
\caption{Hyper-parameters chosen for each neural network model}
\label{tab:hyper}
\begin{tabular}{|c|c|c|c|c|c|c|}
\hline
 & \multicolumn{2}{l|}{Seq2Seq} & Seq2AST & Inst2AST & Inst2AST  & EP \\ \hline
Batch Size & \multicolumn{2}{c|}{50} & 50 & 50 & 100 & 10 \\ \hline
Number of RNN layer & \multicolumn{2}{c|}{2} & 1 & 1& 1 & 1 \\ \hline
Encoder RNN cell & \multicolumn{2}{c|}{LSTM} & LSTM & N-ary LSTM & N-ary LSTM & - \\ \hline
Decoder RNN cell & \multicolumn{2}{c|}{LSTM} & Tree LSTM & LSTM &Tree LSTM & GRU \\ \hline
Learning rate & \multicolumn{6}{c|}{\begin{tabular}[c]{@{}c@{}}Decay the learning rate by a factor of 0.9× when the\\ validation loss does not decrease for 200 mini-batches\end{tabular}} \\ \hline
Hidden state size & \multicolumn{6}{c|}{128} \\ \hline
Embedding size & \multicolumn{6}{c|}{128} \\ \hline
Dropout Rate & \multicolumn{6}{c|}{0.5} \\ \hline
Gradient clip threshold & \multicolumn{6}{c|}{1.0} \\ \hline
Weight Initialization & \multicolumn{6}{c|}{Uniform Random from {[}-0.1,0.1{]}} \\ \hline
\end{tabular}
\end{table}

\section{\sys{} Decompilation Evaluation on x86-64 ISA}
\vspace{-1em}

We also evaluate \sys{}'s performance on x86-64 ISA and summarize the results in Table~\ref{tab:token_accuracy_x86}. We compile the code using \code{gcc -O0} configuration. \sys{}'s decompilation token accuracy on x86-64 achieves is on average 6\% lower than the one on MIPS architecture. This is mainly because: (i) x86 has more advanced instructions types that support different granularity of read-write and data movements while MIPS support only 32-bits read/write operations. As such the number of input token types in x86-64 is larger than than the one in MIPS ISA for the same high-level program; (ii) the branch flag is stored in condition register which is not explicit visible as part of the branch instructions as MIPS. Thus, it is hard to reverse the CFG compared to MIPS architecture. 

\vspace{-1em}
\begin{table}[ht!]
\centering
\caption{Token accuracy across benchmarks on x86 assembly input.}
\label{tab:token_accuracy_x86}
\scalebox{1}{
\begin{tabular}{|c|c|c|c|c|c|c|}
\hline
Benchmarks & Seq2Seq+Attn & Seq2AST+Attn & Inst2AST+Attn \\ \hline
Karel$_S$ & 96.73 & 99.50& \textbf{99.61} \\ \hline
Math$_S$ & 90.16 & 96.19 & \textbf{96.50} \\ \hline
NE$_S$ & 85.73 & 88.76 & \textbf{89.33} \\ \hline
(Math+NE)$_S$ & 77.51 & 82.15 & \textbf{87.84} \\ \hline
Karel$_L$ & 95.20 & 96.17 & \textbf{96.41} \\ \hline
Math$_L$ & 86.64 & 91.55 & \textbf{92.60} \\ \hline
NE$_L$ & 78.56 & 80.63 & \textbf{83.19} \\ \hline
(Math+NE)$_L$ & 73.64 & 77.67 & \textbf{81.12} \\ \hline
\end{tabular}
}
\end{table}

\newpage

\section{Examples Benchmarks Task}
We present the examples in Figure~\ref{lst:e1_bench} and~\ref{lst:e2_bench}. For simplicity, we list only the code snippet example for each benchmarks. The training dataset has different program length and variable numbers. For \code{Math+NE}, the dataset is build by replacing the variables in NE with functions. 

\renewcommand{\lstlistingname}{Figure}
\renewcommand{\lstlistlistingname}{List of \lstlistingname s}

\definecolor{codegreen}{rgb}{0,0.6,0}
\definecolor{codegray}{rgb}{0.5,0.5,0.5}
\definecolor{codepurple}{rgb}{0.58,0,0.82}
\definecolor{backcolour}{rgb}{0.95,0.95,0.92}
 
\lstdefinestyle{mystyle}{
    commentstyle=\color{codegreen},
    keywordstyle=\color{magenta},
    numberstyle=\tiny\color{codegray},
    stringstyle=\color{codepurple},
    basicstyle=\footnotesize,
    tabsize=2
}
\lstset{style=mystyle}
\vspace{-2em}
\begin{minipage}{\textwidth}
\begin{parcolumns}{2}
\colchunk{\begin{lstlisting}[style=mystyle,label=lst:e1_bench,language=C++,basicstyle=\footnotesize,caption=Benchmark examples for (i) Pytorch C++ API  (ii) Hacker's Delight]{Name}
// (i) NN construction for MNIST
struct nn::Module {
    Net()
      : conv1(Conv2dOptions(v1,v2,v3)),
        conv2(Conv2dOptions(v2,v4,v3)),
        fc1(v5, v6),
        fc2(v6, v7)
    }
    Tensor forward(Tensor x) {
    x = conv1->forward(x)
    x = torch::max_pool2d(x,2);
    x = torch::relu(x);
    x = conv2->forward(x);
    x = torch::max_pool2d(x,2);
    x = x.view({-1, v5});
    x = torch::relu(x, 2));
    x = fc1->forward(x);
    x = torch::relu(x);
    x = torch::dropout(x);
    x = fc2->forward(x);
    x = torch::log_softmax(x,1)
    return x;
  }
};

// (ii) Hacker's Delight example:
int32_t p25(int32_t x, int32_t y,\ 
int32_t base, int shift) {
  uint32_t o1 = x & base;
  int32_t o2 = x >> shift;
  uint32_t o3 = y & base;
  int32_t o4 = y >> shift;
  uint32_t o5 = o1 * o3;
  int32_t o6 = o2 * o3;
  int32_t o7 = o1 * o4;
  int32_t o8 = o2 * o4;
  int32_t o9 = o5 >> shift;
  int32_t o10 = o6 + o9;
  int32_t o11 = o10 & base;
  int32_t o12 = o10 >> shift;
  int32_t o13 = o7 + o11;
  int32_t o14 = o13 >> shift;
  int32_t o15 = o14 + o12;
  return o15 + o8;
}

\end{lstlisting}}
    \colchunk{\begin{lstlisting}[style=mystyle,label=lst:e2_bench,language=C++,basicstyle=\footnotesize,caption=Benchmark examples for (i) \code{Karel}  (ii) \code{Math} (iii) \code{NE}. (iv) \code{Math+NE}.]{Name}
// (i) Karel
int main(){
    TurnOn();
    TurnOff();
    while(leftIsClear){
    
        PutBeeper();
        TurnLeft();
        if(notFacingNorth){
            continue;
        }
        PickBeeper();
        Move();
    }
    PickBeeper();
}
// (ii) Math
int func(int a, double b, int c, \
double d, double e){
    b=log(a);
    while(islessequal(d,a)){
        e=isgreaterequal(c,b);
        a=cos(e);
    }
    d=atan2(c,i);
    b=atan2(d,e);
    a=fmin(b,c);
}
// (iii) Normal Expressions
int func(int a, int b, double c) {
    a=a-b;
    c=b+c>>a;
    if((c>a)||(b<=e)){
        c=(c+a)/d;
        b=d*c-b+e;
        b=c<<a;
    }
}
// (iv) Math+Normal Expressions
int func(double a, int b,  \ 
double c, double d, double e){
  b=log(a)-atan(d);
  while(isgreater(d,a)||isless(e,d)){
    e=isgreaterequal(c,b);
    a=cos(e);
  }
  d=atan2(c,i);
  b=(atan2(d,e)-fmax(c,a))/ceil(c);
  a=fmin(b,c)*asin(d)/pow(d,f);
}
\end{lstlisting}}
\colplacechunks
\end{parcolumns}
\end{minipage}

\newpage
\section{Examples of Traditional Decompiler Results}
\vspace{-1em}
We present the decompiled results of traditional decompiler and \sys{} in Figure~\ref{lst:e1} and~\ref{lst:e2}. Note that RetDec is the most recent published decompiler with more than 500K lines of code, which is 100$\times$ larger than \sys{}. Their toolkit size is $\sim$5GB which is around 500$\times$ larger than the size of \sys{}'s neural networks($\sim$10MB).

\renewcommand{\lstlistingname}{Figure}
\renewcommand{\lstlistlistingname}{List of \lstlistingname s}

\vspace{-2.3em}
\begin{minipage}{\textwidth}
\begin{parcolumns}{2}
\colchunk{\begin{lstlisting}[style=mystyle,label=lst:e1,language=C,basicstyle=\footnotesize,caption=Source Code and corresponding decompiled results from (i) \sys{} and (ii) RetDec. This example shows that the state-of-the-art decompiler fails to preserve the functionality and semantics. \code{g1 to g7} are global variable that is used to pass parameters.]{Name}
// source code
int code(double a, double b,\ 
double c){
    b=log(c);   
    while(b<a){
        a=cos(b)*c;
    }
    return b;
}

// (i) decompiled code from Coda
int func(double v1, double v2, \
double v3){
    v2=log(v3);
    while(v2<v3){
        v1=cos(v2)*v3;
    }
    return v2;
}
// (ii) decompiled code from RetDec
int32_t code(void) {
    log((float64_t)(int64_t)g1);
    float64_t v1 = g6;
    if (g7 < v1) {
        float64_t v2 = v1 * v1; 
        float64_t v3 = v1 * v2; 
        float64_t v4 = v2 + v3; 
        while (v3 < v4) {
            v1 = v4;
            v2 = v1 * v1;
            v3 = v1 * v2;
            v4 = v2 + v3;
        }
        g6 = v4;
    }
    __asm_cfc1(g8);
    int32_t v5 = __asm_cfc1(g8);
    __asm_ctc1(v5 & -4 | 1, g8);
    __asm_ctc1(v5, g8);
    return (float32_t)g6;    
}
\end{lstlisting}}
\colchunk{\begin{lstlisting}[label=lst:e2,language=C,basicstyle=\footnotesize,caption=Source Code and corresponding decompiled results from (i) \sys{} and (ii) RetDec. The result shows that RetDec succeeds in preserving the functionality of the code in this case while the recovered high-level code is still difficult for human understanding.]{Name}
// source code
int main(int argc, char *argv[]){
    int a = atoi(argv[1]);;
    int b = atoi(argv[2]);;
    int c = atoi(argv[3]);;
    a = b * c - 1 ;
    if( a > 1 ){
        a = b + c;
        c = a * c - b;
    }
    return c;
}
// (i) decompiled code from Coda
int main(int argc, char *argv[]){
    int v1 = atoi(argv[1]);;
    int v2 = atoi(argv[2]);;
    int v3 = atoi(argv[3]);;
    v1 = v2 * b3 - 1 ;
    if( v1 > 1 ){
        v1 = v2 + v3;
        v3 = v1 * v3 - v2;
    }
    return c;
}
// (ii) decompiled code from RetDec
int main(int argc, char ** argv) {
  int32_t v1 = (int32_t)argv; 
  atoi((char*)*(int32_t*)(v1 + 4));
  int32_t v2 = *(int32_t*)(v1 + 8); 
  int32_t v3 = *(int32_t*)(v1 +12);
  int32_t result;
  if (v3 * v2 >= 3) {
    result = (v3 + v2) * v3 - v2;
  } else {
    result = v3 * v2 < 3;
  }
  return result;
}
\end{lstlisting}}
\colplacechunks
\end{parcolumns}
\end{minipage}